\documentclass[12pt]{article}
\usepackage[english]{babel}
\usepackage{url}
\usepackage[utf8x]{inputenc}
\usepackage{amsmath}
\usepackage{graphicx}
\graphicspath{{images/}}
\usepackage{parskip}
\usepackage{fancyhdr}
\usepackage{hyperref}
\usepackage{vmargin}
\usepackage{listings} 
\usepackage[font=small,skip=0pt]{caption}
\setmarginsrb{3 cm}{2.5 cm}{3 cm}{2.5 cm}{1 cm}{1.5 cm}{1 cm}{1.5 cm}

\title{Ring Oscillator and its application as Physical Unclonable Function (PUF) for Password Management}								
\author{Alireza Shamsoshoara}								
\date{\today}											

\makeatletter
\let\thetitle\@title
\let\theauthor\@author
\let\thedate\@date
\makeatother

\begin{document}


\begin{titlepage}
	\centering
    \vspace*{0.5 cm}
	\rule{\linewidth}{0.2 mm} \\[0.4 cm]
	{ \huge \bfseries \thetitle}\\
	\rule{\linewidth}{0.2 mm} \\[1 cm]
	
	\begin{minipage}{0.4\textwidth}
		\begin{flushleft} \large
			\emph{Author:}\\
			\theauthor
			\end{flushleft}
			\end{minipage}~
			\begin{minipage}{0.4\textwidth}
			\begin{flushright} \large
			\emph{} \\
		\end{flushright}
	\end{minipage}\\[1 cm]
	
	{\large January, 2019}\\[1 cm]
 
	\vfill
	
\end{titlepage}


\tableofcontents
\pagebreak


\section{Physical Unclonable Function (PUF)}

Mobile and embedded devices are becoming inevitable parts of our daily routine. Similar to other electronic devices such as read access memory (RAM) and storage, mobile devices require to authenticate and to be authenticated in a secure way. Usually, this can be accomplished by servers which possess private information for all devices. Since these devices are inherently mobile and operating in untrusted environments, they are prone to be accessed by untrustworthy users.
One way to address this issue is to plant a secret key in the electrically erasable programmable read-only memory (EEPROM) or battery backed static random-access memory (SRAM) of the device. However, this solution is power consuming and expensive. Moreover, these memories are vulnerable to malicious attacks.

Physical unclonable function (PUF) is a unique physical feature of semiconductor device such as microprocessor that can be generated from the physical conditions such as supply voltage, temperature, etc. The main advantages of PUFs are easy to extract digital fingerprints, inexpensive, and do not require nonvolatile memories such as EEPROM or Non-volatile (nvSRAM).

Physical unclonable functions (PUFs) can be used in order to authenticate or store the secret keys \cite{gassend2002silicon,ruhrmair2012security}. The fundamental characteristic of PUF is that there is not any storage for keeping the secret key. Instead, the PUF extracts the key from the features and attributes of the device. Temperature, humidity, and air pressure affect attributes for each PUF. For instance, Herder and et. al examine gate delay as a feature for deriving the secret key \cite{herder2014physical, shamsoshoara2019overview}. Using physical characteristic in PUFs is advantageous because developing circuits for each PUF is simpler; also, passive attacks are not efficient due to the fact that the secret key is obtained based on physical characteristics when the device is powered on. In addition, nonvolatile memories are more expensive in fabricating and using additional layers during the design. Furthermore, PUFs do not need any external source or battery resource; however, RAMs like SRAM or are in need of some external energy source to store the secret key.

One of the key features of PUF is that, even though the building methods are the same for different PUFs, each PUF acts differently in generating the secret key \cite{herder2014physical}. This concept is called “unique objects” \cite{kirovski2010anti}. For instance, this word is defined as a fingerprint for a device which is assumed as stable and robust over time \cite{ruhrmair2012security}.

\subsection{Methods to design PUFs}

PUFs are categorized based on their applications. 1) low-cost authentication and 2) secure key generation are two main applications for PUFs. Based on these, “Strong PUF” and “Weak PUF” are delineated. In most scenarios, strong PUFs are utilized for authentication and weak PUFs are employed for key storage.
Pappu et al. in 2002 implemented the first strong PUF which was called “physical one-way function” \cite{pappu2002physical}. Also, this PUF was known as optical PUF. It was fabricated based on three main parts 1) a laser point on the XY plane which can point to Z axis; 2) a stationary medium which scatters the beam of the laser; and 3) a recorder device to capture the output of the stationary medium. Another example of strong PUFs is silicon implementation which was introduced by Gassend et al. in 2002 \cite{gassend2002silicon}. In this sample, gate delay is utilized as the source for unclonable randomness. In this method, two multiplexers are considered for each stage and based on different inputs in each stage, a new delay will be rendered.

Suh et al. in 2007 introduced a “ring-oscillator” PUF based on gate delay propagation \cite{suh2007physical}. Ring oscillators are the principal parts of this PUF that are synthesized on a field-programmable gate array (FPGA) or an application-specific integrated circuit (ASIC) \cite{suh2007physical}. This PUF is categorized as the weak PUF since the number of challenges that can trigger the PUF is limited. In this PUF, frequency of ring oscillator determines the output for PUF. Thus, environmental parameters and noise will affect the performance of PUF, and hence error correction is required.
Layman et al. in 2002 introduced the scheme and design of SRAM PUF {layman2004electronic} and SU et al. implemented the first design in 2007 \cite{su20071}. Also, this PUF is classified as weak PUF. In this PUF, a positive feedback loop is utilized to compel the cells of memory into the binary state (saving 0 or 1).

PUFs can be shaped as a black-box challenge-response system. It means that, similar to a function, there is an input for the PUF which is called challenge c and based on the function f(c), response r will be calculated. In this black-box, f() explains the relationship between challenge and response. The domain of f() and the number of challenges that each PUF can process are the leading difference between strong and weak PUFs. For instance, the number of challenges that weak PUF can process is limited, on the other hand, strong PUF is capable of supporting a large number of challenge and response pairs (CRPs) \cite{herder2014physical}.

\subsubsection{Ring Oscillator}

\begin{figure*}[ht]
	\centering
	\includegraphics[width=0.4\textwidth,keepaspectratio]{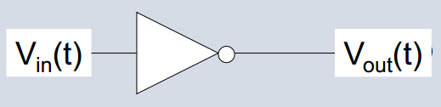}
	\caption{Simple NOT Gate}
	\label{fig:Fig1}
\end{figure*}

\begin{figure*}[ht]
	\centering
	\includegraphics[width=0.6\textwidth,keepaspectratio]{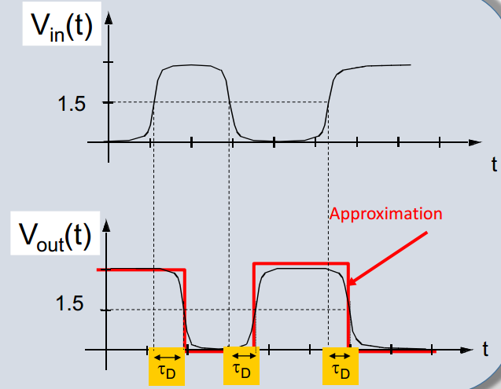}
	\caption{Input and Output Voltage of an Inverter}
	\label{fig:Fig2}
\end{figure*}

A ring oscillator is a device composed of an odd number of NOT gates in a ring, whose output oscillates between two voltage levels, representing true and false. The NOT gates, or inverters are attached in a chain and the output of the last inverter is fed back into the first one. Figure \ref{fig:Fig1} and \ref{fig:Fig2} demonstrate the input and out voltage of a single inverter.

\begin{figure*}[ht]
	\centering
	\includegraphics[width=0.9\textwidth,keepaspectratio]{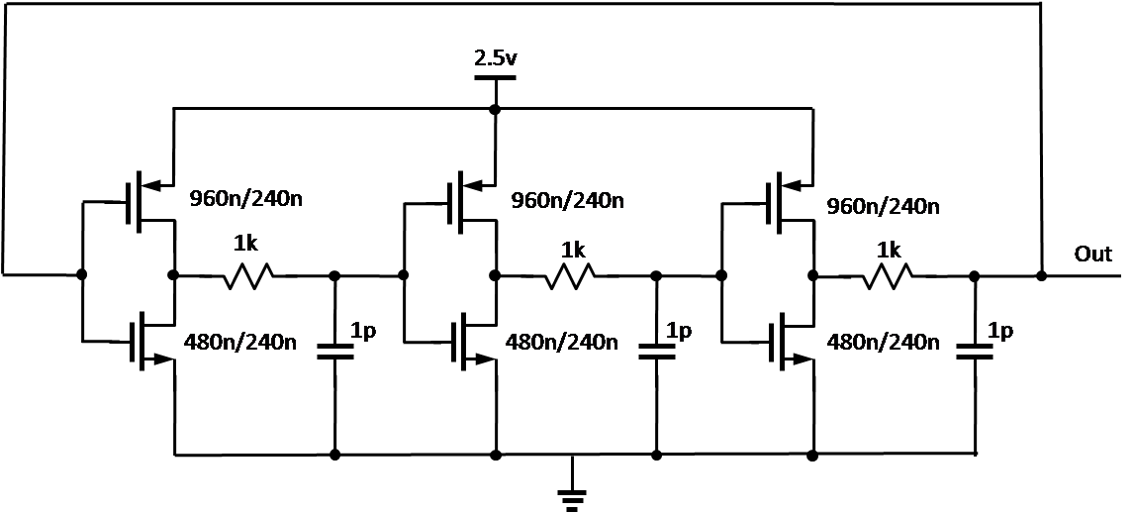}
	\caption{Transistor level of an inverter}
	\label{fig:Fig4}
\end{figure*}

Figure \ref{fig:Fig4} shows the transistor level of a ring oscillator with three inverter. Each gate is attached to one resistor and one capacitor. We can assume that these discrete components are the instinct behaviors of the transistor. PMOS and NMOS transistor are considered for this structure of each inverter.

\subsubsection{Frequency of Ring Oscillator}

The frequency of the ring oscillator is related to the values of resistor and capacitor of each line, If we assume that, all stages have same value and the number of NOT gates is equal to n, the frequency of the output is equal to:

\begin{align}
&f(Hz) = \frac{1}{2 * \tau_{D} * n} 
\\
\nonumber
&= \frac{1}{1.38 * R * C * n}
\end{align}

Where $\tau_{D}$ is $0.69RC$.

\subsubsection{Simulation of Ring Oscillator}

\begin{figure*}[ht]
	\centering
	\includegraphics[width=0.9\textwidth,keepaspectratio]{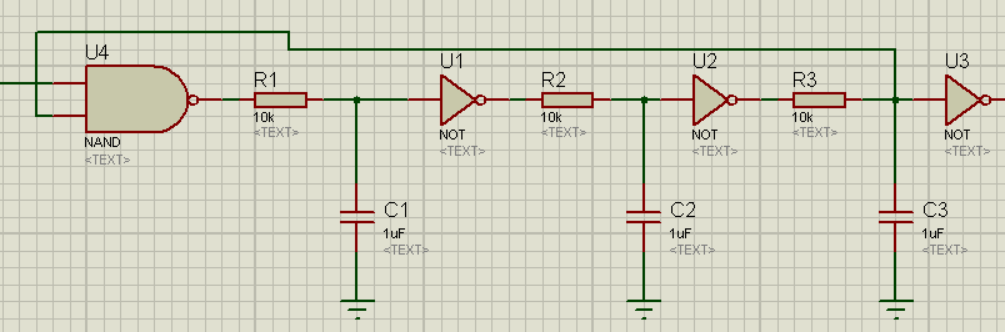}
	\caption{One line of Ring Oscillator}
	\label{fig:Fig3}
\end{figure*}

For designing the ring oscillator, I considered NAND and NOT gates to produce the frequency of the each line. Figure \ref{fig:Fig3} demonstrates more details of my simulated ring. In this example, three NOT gates and one NAND gate are considered as components of this circuit. For simulating the output of this circuit, I used an electrical engineering software which is called Proteus (LAB CENTER). In this application, all of the elements are considered as ideal, so to considered the gate delay of elements, discrete resistors and capacitors are considered in order to simulate the charge and discharge time of each gate.   

This ring oscillator has two inputs and one output, At first the first input in attached to the GND signal and the output of the second inverter is plugged to the second input of the NAND gate. When the input is switched to high voltage from the ground, the capacitors of other stages start charging and discharging and feed back is fed to other input of the NAND gate.

\begin{figure*}[ht]
	\centering
	\includegraphics[width=1.0\textwidth,keepaspectratio]{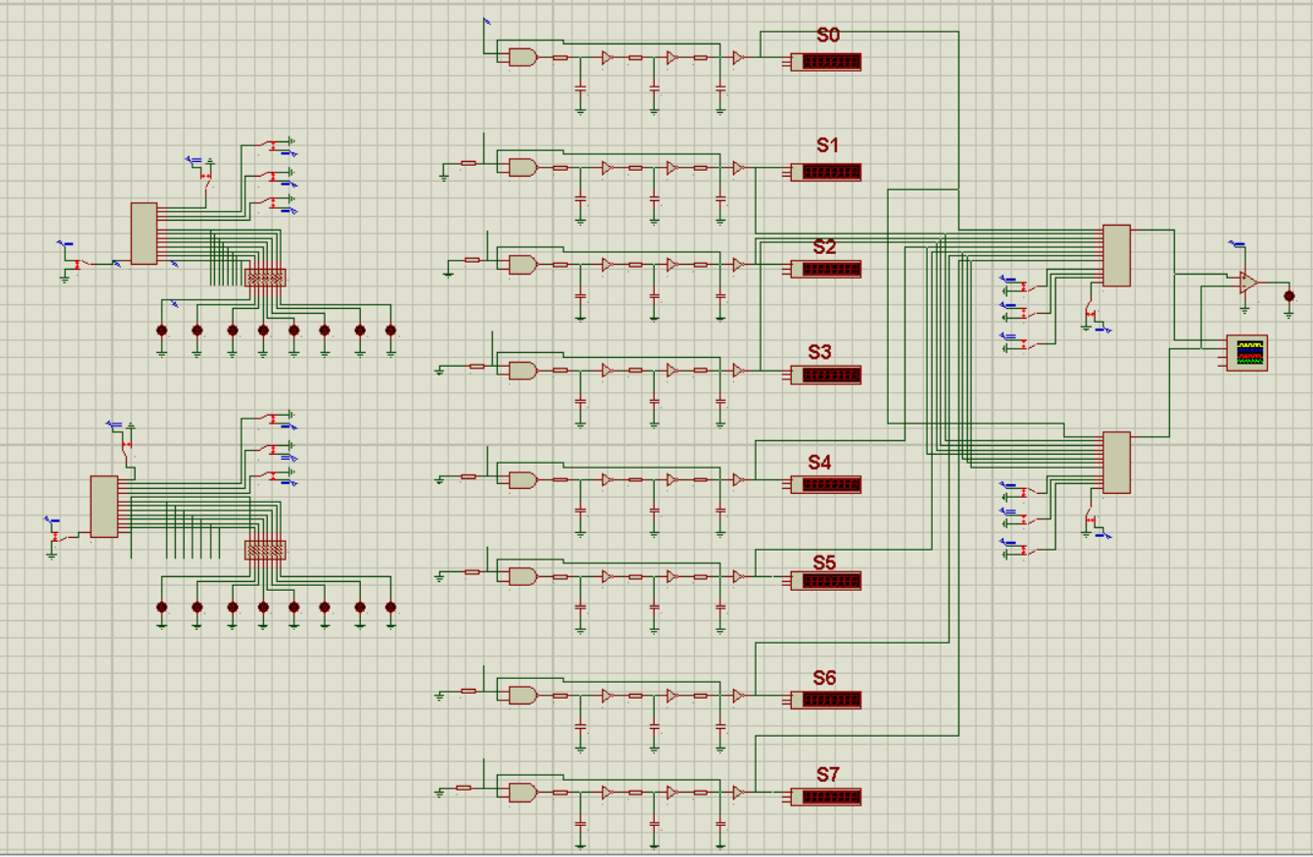}
	\caption{Simulation of the Ring Oscillator in Proteus}
	\label{fig:Figproteus}
\end{figure*}

For the rest of the simulation, 8 lines are considered for the whole circuit which each lines has three NOT and one NAND gate. Two multiplexers are considered for the input selection of the lines. Each multiplexer is 3 to 8, hence, it has One input, 8 outputs, and three selectors. So by choosing only three bits for each element of pair we can choose a ring line to start oscillating. Also other two de-multiplexer are placed in the output with same structure as input to extract the proper lines frequency. Finally two outputs are attached to an oscilloscope for visualization. Figure \ref{fig:Figproteus} illustrates the details of this simulation and all components.

\begin{figure*}[ht]
	\centering
	\includegraphics[width=1.0\textwidth,keepaspectratio]{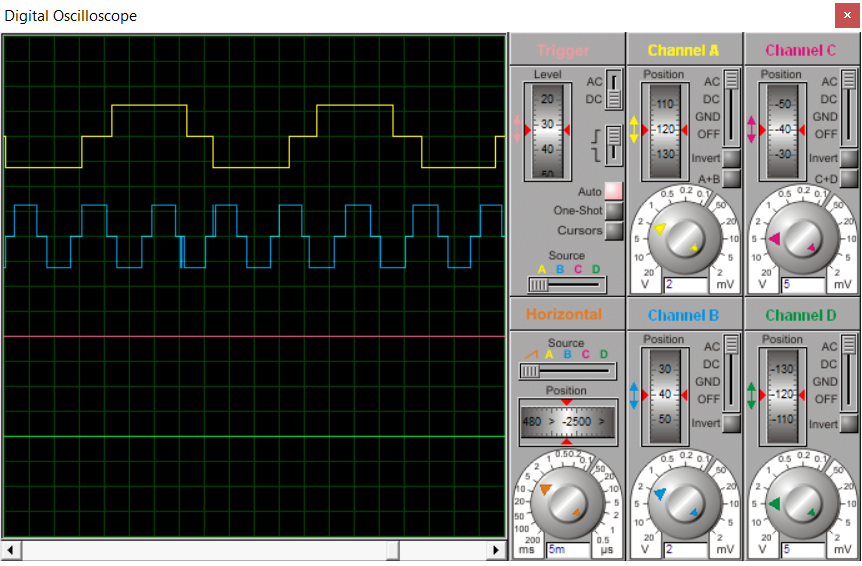}
	\caption{Oscillation of lines 5 and 7 in figure \ref{fig:Figproteus}}
	\label{fig:oscope}
\end{figure*}

Also, Figure \ref{fig:oscope} shows two output (line 5 and 7) of the simulated circuit. Because of different values of capacitors these two frequencies are different with each other.

\subsubsection{Implementation of Ring Oscillator}
\begin{figure*}[ht]
	\centering
	\includegraphics[width=1.0\textwidth,keepaspectratio]{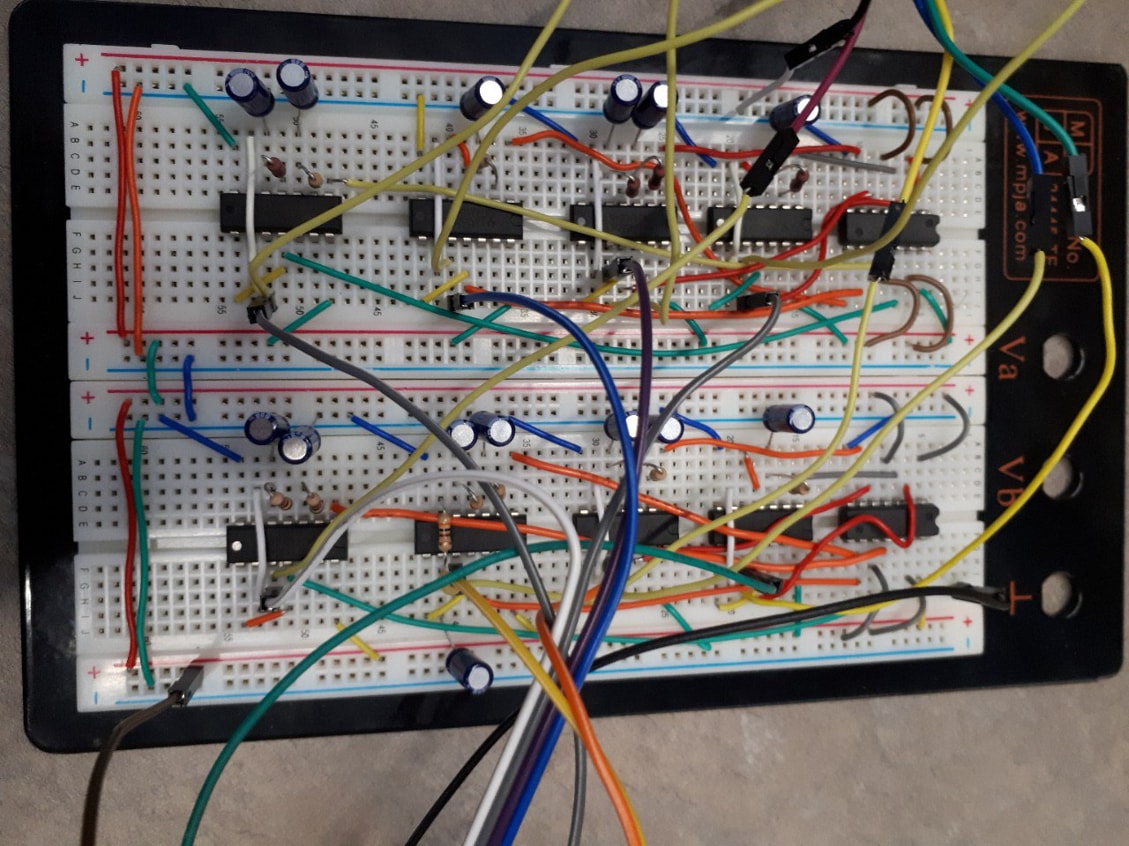}
	\caption{Implemented Circuit for the ring Oscillator}
	\label{fig:Circuit}
\end{figure*}

In order to implement the ring oscillator, CMOS ICs such as 74HC04 and 74HC00 are placed on breadboard for NOT and NAND gates. Again, we used discrete resistors and capacitors in the circuit. The reason is that, without using any resistor or capacitor the frequency is too high to measure that with microcontroller that I'm suing for this project. Also, because I couldn't find any mutliplexer or de-multiplexer, I attached all of the inputs of rings to the VCC and with aid of the microcontroller I measure the frequency of each output simultaneously.  Figure \ref{fig:Circuit} demonstrates the breadboard and other elements on that.

\begin{figure*}[ht]
	\centering
	\includegraphics[width=0.9\textwidth,keepaspectratio]{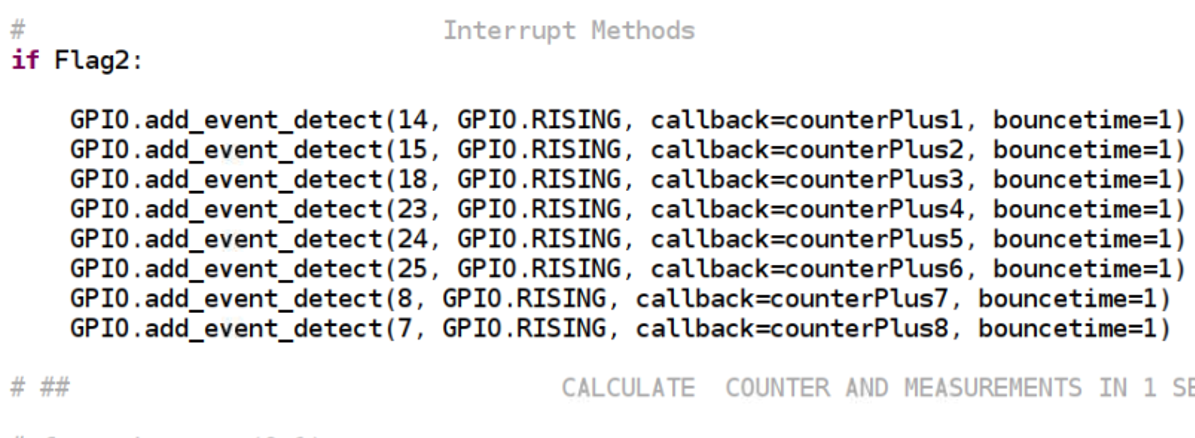}
	\caption{Interrupt structure}
	\label{fig:interrup}
\end{figure*}

\begin{figure*}[ht]
	\centering
	\includegraphics[width=0.5\textwidth,keepaspectratio]{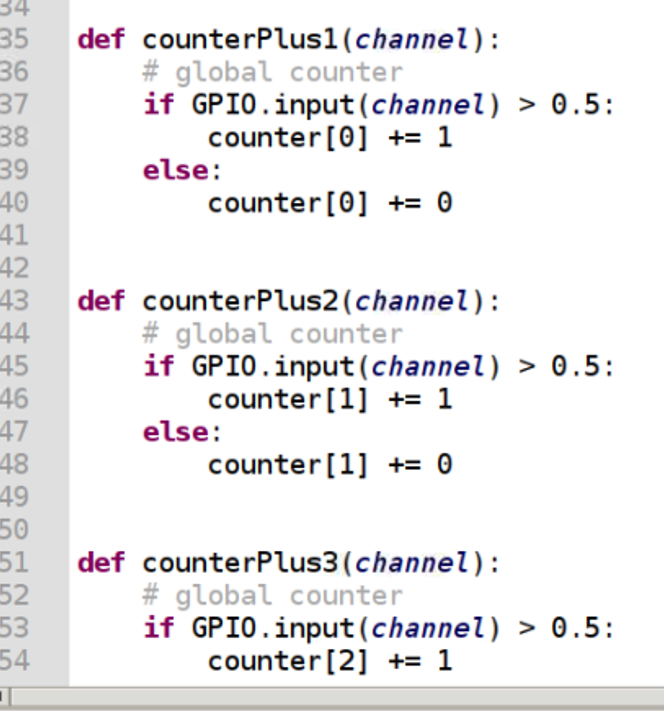}
	\caption{callback function}
	\label{fig:callback}
\end{figure*}

For measuring the frequency, raspberry pi is used for the microcontroller purpose. Interrupts and timer of micro are used in raspberry pi to detect the edge of signals and count the pulse in a definite amount of time which is 0.5 second in this project. Each interrupt is enabled based on the occurred event on GPIO pins of the micro and after edge detection, a callback function is called to count the pulse for the proper outputs. The structures of interrupts and callback functions are depicted in Figure \ref{fig:interrup} and \ref{fig:callback}.

\begin{figure*}[ht]
	\centering
	\includegraphics[width=0.8\textwidth,keepaspectratio]{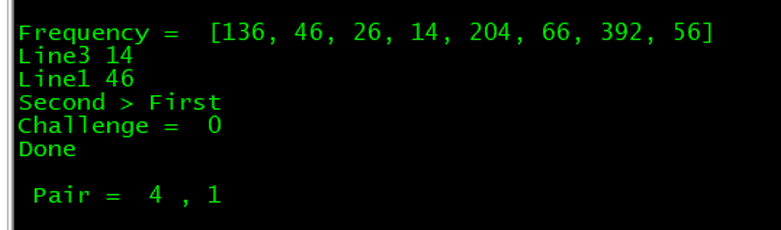}
	\caption{Measured Frequency in Raspberry Pi}
	\label{fig:rpifreq}
\end{figure*}

The whole code of the python language is attached in appendix section. After measuring the frequency of all lines, we found them as $(136, 46, 26, 14, 204, 66, 394, 56)$ which is showed in figure \ref{fig:rpifreq} in raspberry Pi shell.

\begin{figure*}[ht]
  \vspace{0pt}
	\centering
	\includegraphics[width=1\textwidth,keepaspectratio]{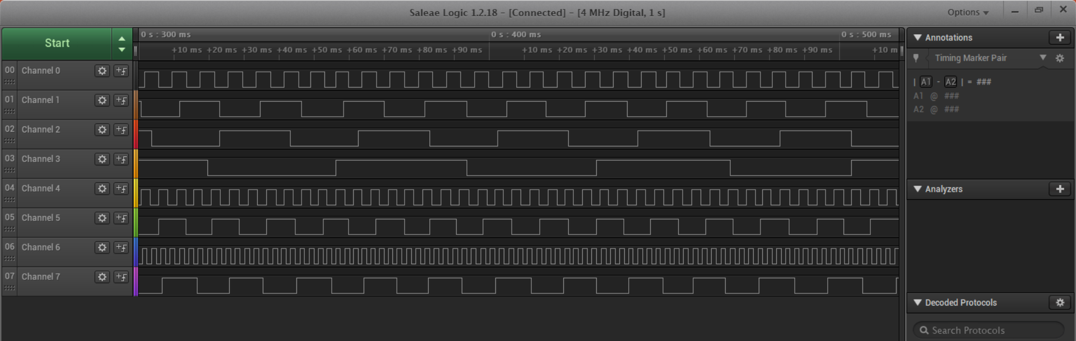}
	\caption{Output signal by logic analyzer}
	\label{fig:logicanalyzer}
\end{figure*}

One of the problem that I had during this project was related to trust of measuring. So, instead of using the oscilloscope I ordered a logic analyzer for comparing the frequency of each line. The price of this device is only 10\$ and link is placed here:

\url{https://www.amazon.com/HiLetgo-Analyzer-Ferrite-Channel-Arduino/dp/B077LSG5P2/ref=sr_1_3?ie=UTF8&qid=1525562576&sr=8-3&keywords=logic+analyzer}

After Comparing the logic analyzer output with the measured frequencies by raspberry pi, I figured out that the values are close to each other and I can trust my frequency counter by the raspberry pi. The output of pulses in the application is depicted in figure \ref{fig:logicanalyzer}. 

\section{Password Management with Ring Oscillator as PUF}

\begin{figure*}[ht]
  \vspace{0pt}
	\centering
	\includegraphics[width=1\textwidth,keepaspectratio]{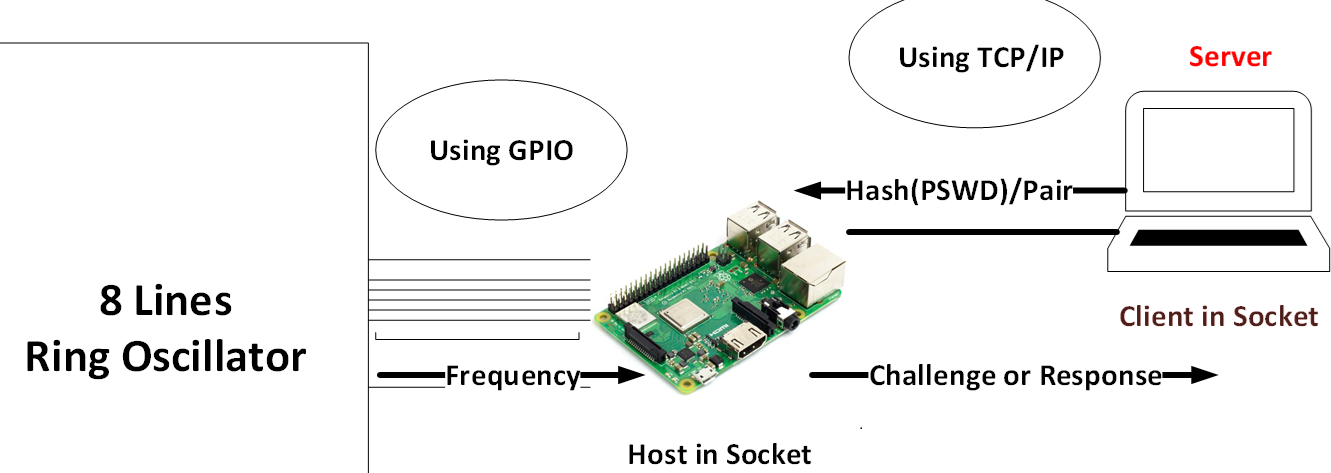}
	\caption{Skeleton of the password management system with raspberry pi and ring oscillator}
	\label{fig:skeleton}
\end{figure*}

Password manager systems are used in order to keep a database of user names and password for clients in order for registration and authentication. This table plays a crucial role for authentication in many systems. Hence, security is one of the important features of these systems. Normally, in these tables only user names and passwords are stored; however, they are vulnerable to many attacks. So, the first method of enhancing the security is using the hash instead of the original data, but, advanced hackers can break these hash data again. Hence, the method that we are going to use in this project is utilizing PUF to store the challenge of each user instead of saving the hash of password. In this project, the computer is examined as the database of password management. Databased is implemented in MATLAB, when a user wants to register for an account based on ID and password, 16 pairs of bits (0 to 7) which is the hash(password)) is sent to the raspberry pi, then raspberry pi based on the requested pair, will calculate the frequencies of the proper lines of oscillator, then compare the value of them and produce the challenge and send it back to the computer. For each user we consider 16 pairs which are equal to 32 characters of hash(password). Then we store the challenge in correct row and column of the table. It's noticeable to mention that row and column of the table is calculated based on the XOR(hash(ID), hash(password)). The table in this project has 16 by 16 cells. So for the sake of collisions, we consider structure table instead of matrix table. Because of this, each element of the structure is assumed as a cell. Hence, when collision happen, we can handle that by just appending new challenge in the same location without being worry about the space complexity of the problem. Figure \ref{fig:skeleton} shows the detail of the skeleton. 

For communication between the raspberry pi and the ring oscillator, we only use GPIO of the micro controller to sense the events of pulses. Communication between the computer and raspberry pi is accomplished by using the TCP/IP protocol. So I put both devices in a same local area network and use socket binding for RPI and the computer. In this scenario, Computer and RPi are assumed as client and the host. The RPi listens to the data which is being sent by the computer then send back the challenge or response. MATLAB code of the password management system is attached to the appendix section.

\section{Demo and Test}
\begin{figure*}
	\centering
	\includegraphics[width=1\textwidth,keepaspectratio]{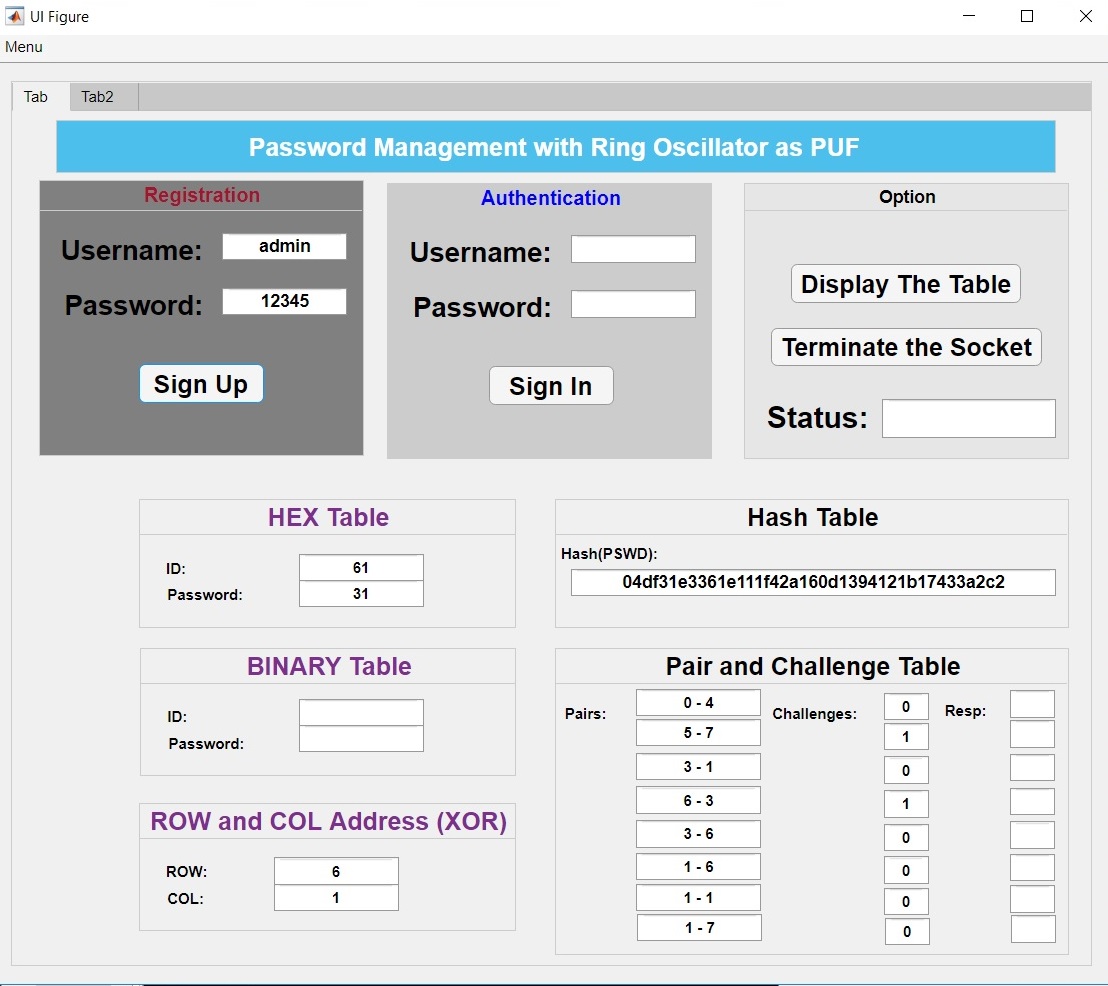}
	\caption{Register an account}
	\label{fig:register}
\end{figure*}

\begin{figure*}
	\centering
	\includegraphics[width=1\textwidth,keepaspectratio]{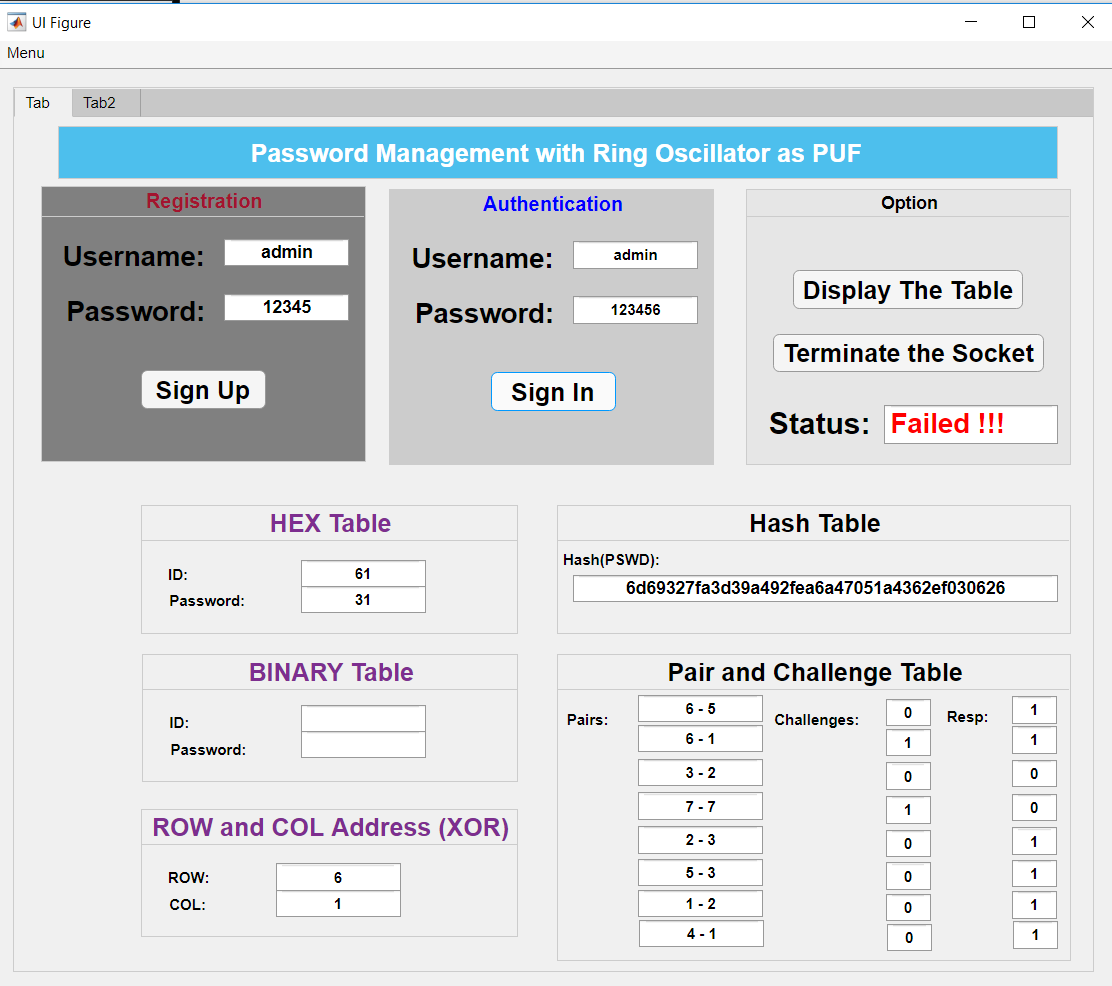}
	\caption{Problem in Authentication}
	\label{fig:failed}
\end{figure*}

\begin{figure*}
	\centering
	\includegraphics[width=1\textwidth,keepaspectratio]{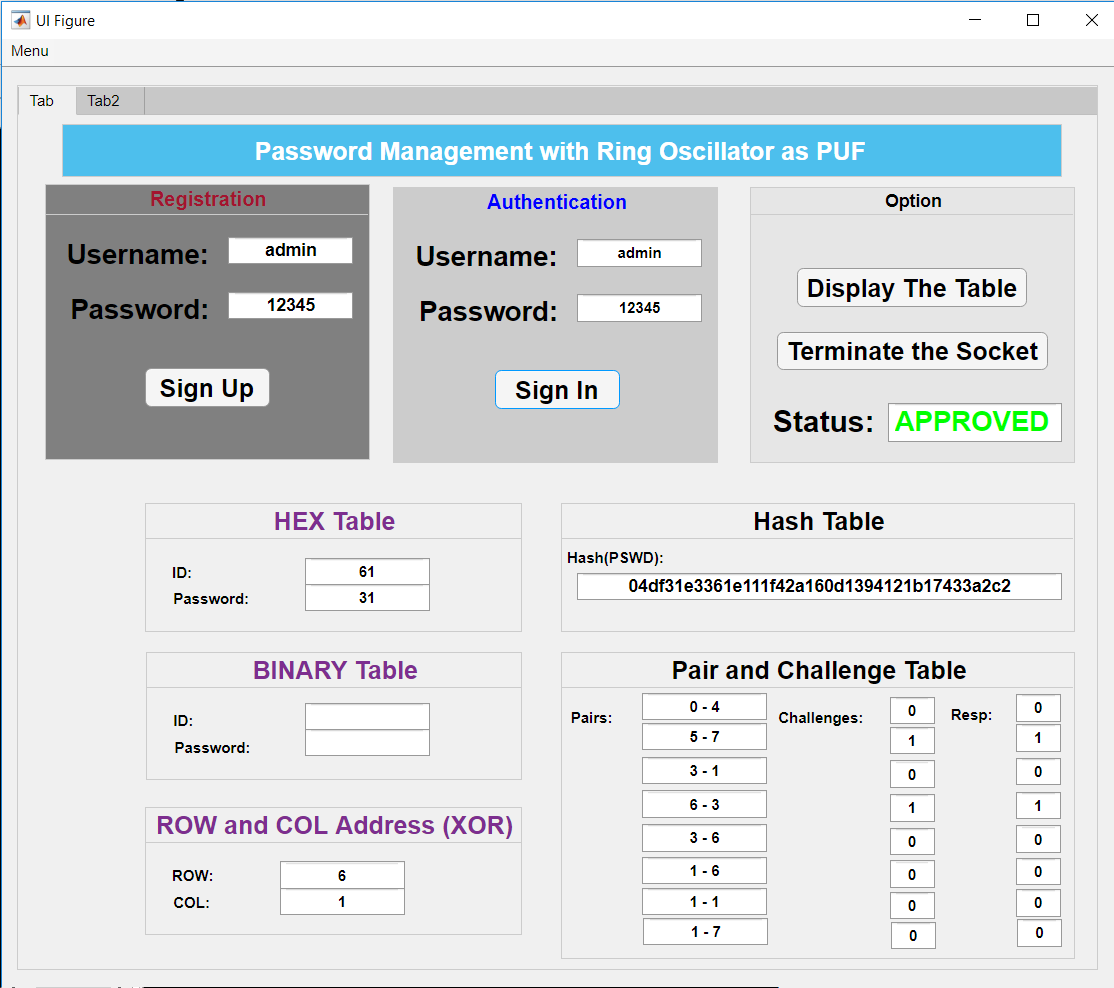}
	\caption{Approved Authentication}
	\label{fig:approved}
\end{figure*}

For this project, a graphical user interface is designed in MATLAB for better sense of the project. There are three modes in this application: 1) Registration, 2) Authentication, and 3) Display the table. Figure \ref{fig:register} shows how a user can register with user name and password and it shows the challenge contents of the first 8 bits. Figure \ref{fig:failed} depicts the failed status when a user enters wrong password, and finally Figure \ref{fig:approved} shows an approved authentication based on correct challenge and response pair.

\newpage

\bibliographystyle{plain}
\bibliography{biblist.bib}

\end{document}